\documentclass[letterpaper]{article} % DO NOT CHANGE THIS
\usepackage{aaai24}  % DO NOT CHANGE THIS
\usepackage{times}  % DO NOT CHANGE THIS
\usepackage{helvet}  % DO NOT CHANGE THIS
\usepackage{courier}  % DO NOT CHANGE THIS
\usepackage[hyphens]{url}  % DO NOT CHANGE THIS
\usepackage{graphicx} % DO NOT CHANGE THIS
\usepackage{natbib}  % DO NOT CHANGE THIS AND DO NOT ADD ANY OPTIONS TO IT
\usepackage{caption} % DO NOT CHANGE THIS AND DO NOT ADD ANY OPTIONS TO IT
\usepackage{algorithm}
\usepackage{algorithmic}
\usepackage{amsmath}
\usepackage{tabularx}
\usepackage{booktabs}
\usepackage{amsfonts}

\usepackage{color}
\usepackage{xcolor}

\newcommand{\tang}[1]{{\color{red}{#1}}}

\usepackage{newfloat}
\usepackage{listings}
\DeclareCaptionStyle{ruled}{labelfont=normalfont,labelsep=colon,strut=off} % DO NOT CHANGE THIS
\floatstyle{ruled}
\newfloat{listing}{tb}{lst}{}
\floatname{listing}{Listing}
\title{GIN-SD: Source Detection in Graphs with Incomplete Nodes via Positional Encoding and Attentive Fusion}
\author{
    Le Cheng\textsuperscript{\rm 1,2}, Peican Zhu\textsuperscript{\rm 2}\thanks{Corresponding author.}, Keke Tang\textsuperscript{\rm 3}, Chao Gao\textsuperscript{\rm 2}, Zhen Wang\textsuperscript{\rm 1,2}\thanks{Corresponding author.}\\
}
\affiliations{
    \textsuperscript{\rm 1}School of Computer Science, Northwestern Polytechnical University (NWPU)\\
    \textsuperscript{\rm 2}School of Artificial Intelligence, Optics and Electronics (iOPEN), Northwestern Polytechnical University (NWPU)\\
    \textsuperscript{\rm 3}Cyberspace Institute of Advanced Technology, Guangzhou University\\
    ericcan@nwpu.edu.cn, w-zhen@nwpu.edu.cn
}

\usepackage{bibentry}
\begin{document}

\maketitle

\begin{abstract}

Source detection in graphs has demonstrated robust efficacy in the domain of rumor source identification. Although recent solutions have enhanced performance by leveraging deep neural networks, they often require complete user data. In this paper, we address a more challenging task, rumor source detection with incomplete user data, and propose a novel framework, i.e., Source Detection in Graphs with Incomplete Nodes via Positional Encoding and Attentive Fusion (GIN-SD), to tackle this challenge. Specifically, our approach utilizes a positional embedding module to distinguish nodes that are incomplete and employs a self-attention mechanism to focus on nodes with greater information transmission capacity. To mitigate the prediction bias caused by the significant disparity between the numbers of source and non-source nodes, we also introduce a class-balancing mechanism. Extensive experiments validate the effectiveness of GIN-SD and its superiority to state-of-the-art methods.

\end{abstract}

\section{Introduction}

% \tang{
Source detection in graphs represents a fundamental challenge in mathematics and plays a vital role in rumor source detection~\cite{shah2011rumors, ling2022source, zhu2022locating, cheng2022path}. Early solutions, such as LPSI~\cite{wang2017multiple}, EPA~\cite{ali2019epa}, and MLE~\cite{pinto2012locating}, primarily rely on source centrality theory~\cite{prakash2012spotting, shah2011rumors} and maximum likelihood estimation in detecting sources. In recent years, with the advancement of deep learning techniques \cite{gao2022novel}, researchers have utilized deep neural networks to encode user attributes and propagation information~\cite{bian2020rumor,wang2022invertible,ling2022source}, significantly refreshing the state-of-the-art records.
% }
%approaches like IVGD \cite{wang2022invertible}, SL-VAE \cite{ling2022source} and Bi-GCN \cite{bian2020rumor} have emerged, taking a deep learning perspective by utilizing user attributes and propagation information as input features, refreshing the state-of-the-art  records.

% To counteract the propagation of malicious information, e.g., the rumors, timely and effective identification of their sources to halt further dissemination is of significant importance \cite{shah2011rumors, ling2022source}. Historically, many methods have primarily relied on the source centrality theory \cite{prakash2012spotting, shah2011rumors} and maximum likelihood estimation \cite{pinto2012locating} for source detection, exemplified by approaches such as LPSI \cite{wang2017multiple}, EPA \cite{ali2019epa}, and MLE \cite{pinto2012locating}. In recent years, with the advent of neural networks, approaches like IVGD \cite{wang2022invertible}, SL-VAE \cite{ling2022source} and Bi-GCN \cite{bian2020rumor} have emerged, taking a deep learning perspective by utilizing user attributes and propagation information as input features, refreshing the state-of-the-art  records.

\begin{figure}[t]
  \centering
  \includegraphics[width=\linewidth]{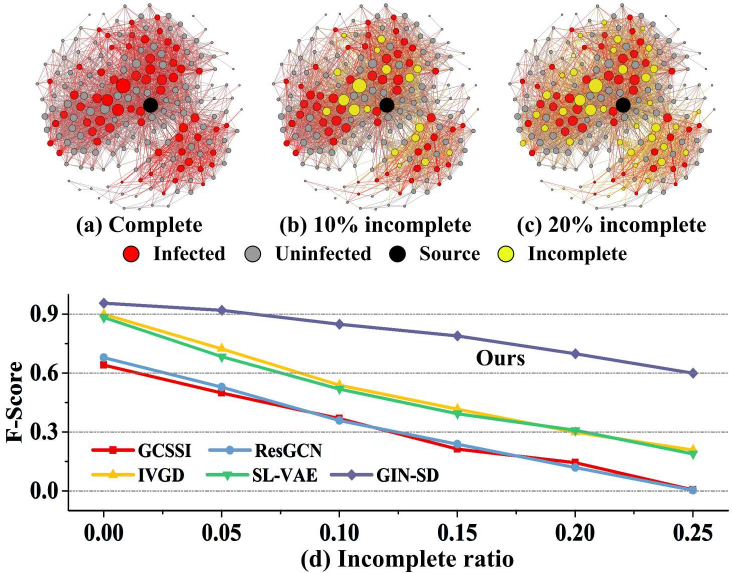}
  \caption{
  Impact of incomplete nodes on source detection: (a-c) graphs with incomplete node ratios of 0\%, 10\%, and 20\%; (d) influence of varying incomplete node ratios on the source detection accuracy for different methods.  As the proportion of incomplete nodes increases, the performance of other methods declines more significantly, while our approach remains less affected.
  %Impact of incomplete nodes on source detection. Panels (a), (b), and (c) respectively depict incomplete node ratios of 0$\%$, 10$\%$, and 20$\%$. Panel (d) illustrates the influence of incomplete node ratios on the source detection accuracy of various methods. As shown, with the increasing proportion of incomplete nodes, the performance of other methods experiences a more pronounced decline, whereas our approach is less affected.
  }
  \label{noise_example}
\end{figure}

\if 0
\begin{figure}[!ht]
  \centering
  \includegraphics[width=\linewidth]{fig_1_v2.pdf}
  \caption{Impact of incomplete nodes on source detection. As the proportion of incomplete nodes increases, the performance of other methods experiences a more pronounced decline, whereas our approach is less affected.}
  \label{noise_example}
\end{figure}
\fi

% \tang{
However, current solutions for source detection are premised on the strict assumption of having access to complete user data, encompassing details such as the forwarding frequency of all users and the time of information reception. Indeed, acquiring such exhaustive user data is exceedingly challenging and sometimes impossible due to time constraints, resource limitations, and privacy protection measures~\cite{du2017community, zhou2019generalized}.
No existing work, to our knowledge, considers the problem of source detection in graph with incomplete nodes. In practice, when user data is incomplete, the majority of cutting-edge solutions falter, as evidenced by the notable performance decline shown in Fig. \ref{noise_example}.
Source detection in graphs with incomplete nodes poses three main challenges.
First, in the process of node information aggregation and transmission, the absent information from incomplete nodes may be erroneously treated as valid data from normal nodes, thus leading to significant feature errors.
Second, since the efficiency of information transmission varies among nodes, e.g., nodes with higher degrees tend to relay information more rapidly, treating all nodes uniformly hinders the training efficiency.
Third, a marked imbalance between the quantities of source and non-source nodes leads the model to favor the non-source set, overlooking the source set, and thus creating a prediction bias.
Intuitively, to handle the above three issues, we should 1) distinguish between incomplete and complete nodes; 2) focus on nodes with superior information transmission capacity; 3) treat source/non-source nodes differently.
% }

\if 0
The primary challenge of source detection under incomplete user data lies in the fact that missing data from certain users may result in misjudgments.
%hinder model convergence and
Existing localized learning-based methods often encounter this predicament, warranting the introduction of learning mechanisms at the global level to facilitate message propagation between nodes.
Moreover, the heterogeneity of users in social networks bestows complexity and stochasticity upon information dissemination \cite{peel2022statistical, zhu2022locating}, thereby emphasizing the significance of prioritizing influential users.
Lastly, given the vast scale of social networks with thousands to millions of users, coupled with the initial dissemination of rumors from only a few sources \cite{dong2022wavefront}, the substantial disparity in the number of users and sources make the model tends to excessively focus on the larger sample set, necessitating the amplification of the weight of the source set while reducing that of the non-source set.
\fi

% \tang{
In this paper, we propose a novel source detection framework in graphs with incomplete nodes (GIN-SD) through positional encoding and attentive fusion.
First, to distinguish incomplete nodes, a positional embedding module is developed to exploit Laplacian Positional Encodings of the infected subgraph, incorporating user states and propagation information into the feature vectors of users.
Second, to focus on nodes with greater information transmission capacity, an attentive fusion module is introduced to employ the self-attention mechanism to automatically allocate varying attention weights to different users.
Finally, to treat source/non-source nodes differently, we introduce a class balancing mechanism that increases the weight of the source set while decreasing the weight of the non-source set, enabling the model to attend to both sets simultaneously.
We validate the effectiveness of our approach on eight publicly available datasets.
Extensive experimental results demonstrate that our approach is robust to missing nodes in a graph, outperforming state-of-the-art methods.
% }

\if 0
In this paper, we propose the GIN-SD: Source Detection of  Graph with Incomplete Node via Positional Encoding and Attentional Fusion. Firstly, to mitigate the impact of lossy information, we embed the Laplacian Positional Encodings (PEs) of the infected subgraph, along with  user states and propagation information, into the feature vectors of users during the Positional Embedding Module (PEM). Subsequently, in the Attentional Fusion Module (AFM), we employ self-attention mechanisms to automatically allocate different attention weights to diverse users. Finally, to tackle the issue of class imbalance in source detection, characterized by a substantial disparity in the number of source and non-source samples, we put forth a class balancing mechanism that augments the weight of the source set while diminishing the weight of the non-source set, enabling the model to simultaneously attend to both sets. In summary, this paper proffers the subsequent contributions:
\fi

Overall, our contribution is summarized as follows:
\begin{itemize}
    \item We are the first to formulate the rumor sources detection under incomplete user data and propose a novel approach to address this issue.
    \item We devise a source detection method of rumors under incomplete user data via positional encoding and attentive fusion mechanism.
    \item We show by experiments that the superiority of the proposed approach in the context of incomplete user data, comparing to baseline methods.
\end{itemize}

\section{Related Work}
\subsection{Infection Status-based Multi-source Detection}
To efficiently address the Multiple Rumor Sources Detection (MRSD) problem, several approaches have been developed. Based on the source centrality theory \cite{prakash2012spotting, shah2011rumors, zhu2017catch}, LPSI selects locally prominent nodes through label propagation without requiring prior information \cite{wang2017multiple}. EPA iteratively calculates the infection time of each node \cite{ali2019epa}. However, these methods do not adequately consider the heterogeneity of users and the stochastic nature of information propagation. Utilizing machine learning techniques, GCNSI \cite{dong2019multiple} and SIGN \cite{li2021propagation} take the states of all users as algorithm inputs, whereas GCSSI focuses on the users infected during the latest wave, known as the wavefront \cite{dong2022wavefront}; from the perspective of model architecture, ResGCN \cite{shah2020finding} incorporates a residual structure that connects GCN layers for message passing. However, these methods fail to consider the randomness of information propagation in heterogeneous networks, and the problem of class imbalance significantly affects the precision of the algorithms. Incorporating the propagation process, IVGD \cite{wang2022invertible} and SL-VAE \cite{ling2022source} introduce diffusion learning mechanisms that thoroughly consider the heterogeneity of users and the stochasticity of information propagation. It's undoubt that obtaining detailed information poses significant challenges due to cost constraints and privacy concerns. Moreover, all the aforementioned methods heavily rely on network snapshot information, assuming the availability of information for all users. However, obtaining a complete network snapshot is immensely challenging due to time constraints, cost limitations, and privacy considerations \cite{du2017community}.

\begin{figure*}[t]
  \centering
  \includegraphics[width=\linewidth]{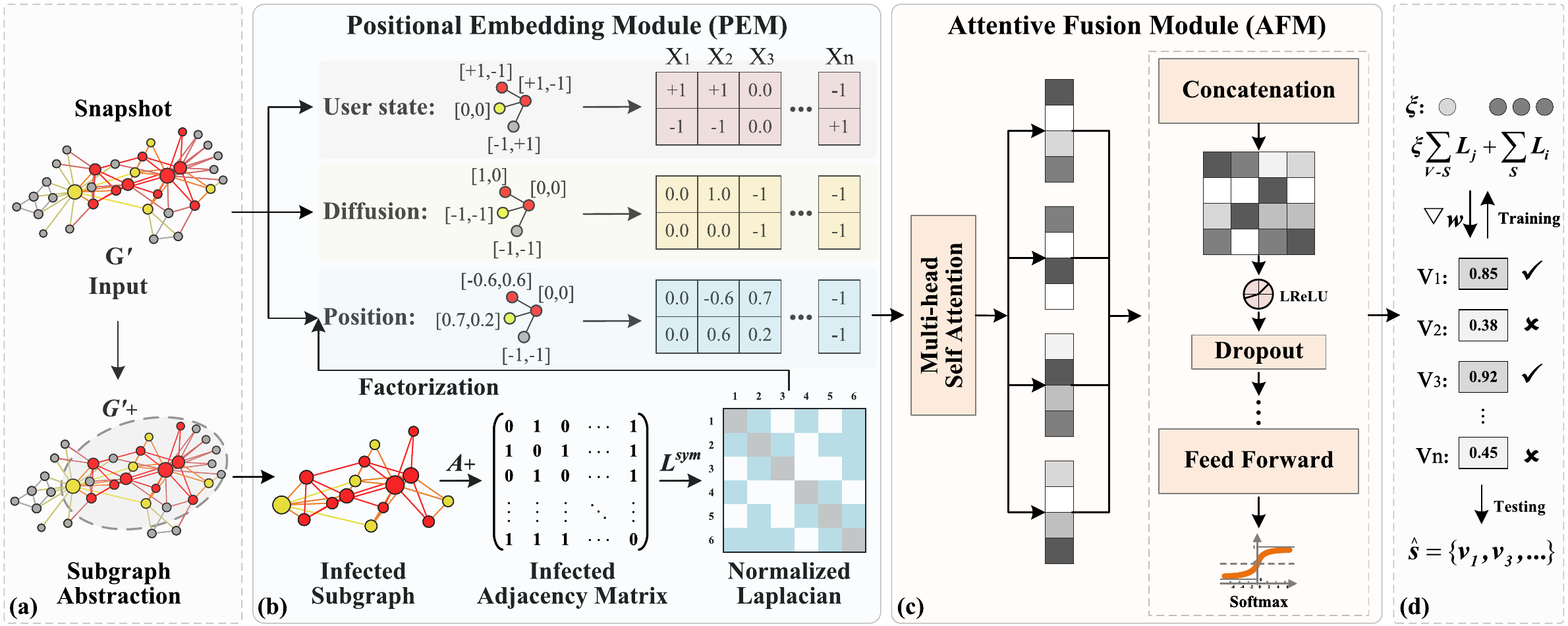}
  \caption{
  Illustration of GIN-SD. (a) The network snapshot $G'$ serves as the input of GIN-SD. (b) The Positional Embedding Module (PEM), where node positional information, along with state and propagation information, is embedded into feature vectors. It is noteworthy that during the position embedding process, the infected subgraph is initially extracted from the acquired snapshot, and the adjacency matrix of the infected subgraph is obtained. Subsequently, the symmetric normalized Laplacian matrix is calculated, and the positional encoding of each node are derived through factorization. (c) The Attentive Fusion Module (AFM) learns node representations through self-attention mechanisms. (d) The training loss is computed using the class-balancing mechanism, and the detected source set $\hat{s}$ is output during the testing phase.
  }
  \label{framework}
\end{figure*}

\subsection{Positional Encodings and Attentive Mechanisms}
The introduction of Graph Neural Networks (GNNs) has enabled the direct application of neural networks, previously designed for Euclidean space, to be applied to graphs (non-Euclidean space) \cite{scarselli2008graph}. The advent of Graph Convolutional Networks (GCNs) has further expedited the advancement of machine learning methods on graphs \cite{kipf2016semi}. GNNs and GCNs effectively learn node representations by leveraging information from the nodes themselves and their neighboring nodes. Moreover, Graph Attention Networks (GAT) empower nodes to allocate distinct attention weights to different neighbors through a multi-head attention mechanism \cite{velivckovic2017graph}. In fact, the models above learn structural node information with invariant node positions \cite{srinivasan2019equivalence}.

In recent years, the Transformer, originally proposed for Natural Language Processing (NLP), has introduced Positional Encodings (PEs) for individual words \cite{han2021transformer}. Which ensures the uniqueness of each word while preserving distance information. Recognizing the merits of global learning based on PEs, PEs learning based on GNNs has also emerged \cite{you2019position, srinivasan2019equivalence, dwivedi2020benchmarking}. For instance, Dwivedi et al. \cite{dwivedi2020benchmarking} employed Laplacian eigenvectors \cite{belkin2003laplacian} as PEs for nodes, enhancing the generative of PEs.

Building upon these, GIN-SD focuses on nodes with greater information transmission capacity through a self-attention mechanisms. Additionally, the Laplacian Positional Encodings of the infected subgraph, along with user states and propagation information are embedded into the user feature vectors to distinguish incomplete nodes.

% harnessing the inductive nature of Laplacian positional encodings, we embed it with user features to bolster global learning. Diverging from the use of PEs for the entire graph, we compute Laplacian PEs specifically for the subgraph of infection. This allows GIN-SD to more comprehensively capture the acquired information.

\section{Problem Formulation}
\subsubsection{Preliminary on Social Networks}
The social networks in the physical world can be abstracted as $G$ = $(V, E)$, where the nodes set $V$ = $\{v_1, v_2, \cdots, v_n\}$ represents the users; and the edges set $E$ = $\{(v_i, v_j) \mid v_i,v_j \in V, i \ne j\}$ indicates the relationships between them. Based on $V$ and $E$, the adjacency matrix $\textit{\textbf{A}}$ $(\textit{\textbf{A}}_{ij} \in \{0,1\}^{n \times n})$ of $G$ is defined as:

\begin{equation}
\textbf{\textit{A}}_{i j} = \left\{\begin{array}{ll}
1, & \left(v_{i}, v_{j}\right) \in E \\
0, & \text {otherwise}.
\end{array}\right.
\end{equation}

\subsubsection{Propagation Process of Social Networks}
Given the definition of $G(V, E)$, the propagation process on $G$ can be represented as a time series $\{X(t), t \geq 0\}$, where $X(t)$ denotes the nodes states in $G$ at time $t$. Specifically, the state $X(t)$ consists of two categories: $G_+$ (positive) and $G_-$ (negative).

When $t$ = $0$, only source set is positive, i.e., $\{s\in G_+, t = 0\}$. After the propagation is triggered by sources, positive user $v_i$ decides whether to propagate the information to its neighbors based on individual forwarding probability $p_i$. Different classical models, such as influence-based Independent Cascade (IC) model \cite{wen2017online}, infection-based Susceptible-Infected (SI) \cite{barthelemy2004velocity} and Susceptible-Infected-Recovered (SIR) \cite{parshani2010epidemic} models, are proposed to simulate the aforementioned propagation process.
\vspace{-0.5mm}

\subsubsection{Source Detection in Graphs with Complete Nodes}
As the propagation unfolds and the rumor reaches a certain significance threshold, specifically when $\theta\%$ of the nodes in the network are infected, a network snapshot $G'(T,U,P)$ is obtained includes: 1) network topology $T$; 2) user information $U$: user states, information forwarding frequency; 3) propagation information $P$: reception time of information, information propagator.

Based on the aforementioned definitions, the source detection problem with complete nodes can be formalized as:
\begin{equation}
\hat{s}=f(G'(T,U,P)),
\end{equation}
where $f(\cdot)$ is the corresponding sources detection methodology, and $\hat{s}$ represents the detected source set.
\vspace{-0.5mm}

\subsubsection{Source Detection  in Graphs with Incomplete Nodes}
In practice, source detection with complete nodes is exceedingly challenging and sometimes impossible due to time constraints, resource limitations, and privacy protection solutions \cite{du2017community,zhou2019generalized}. Leading to incomplete user data in $G'$:
\begin{equation}
U'=(1 - \delta) U, P'=(1 - \delta) P,
\end{equation}
$\delta$ represents the incomplete ratio of user data. Hence the source detection in graphs with incomplete nodes is formalized as:
\begin{equation}
\hat{s}=f(G'(T,U',P')),
\end{equation}

\subsubsection{Discussion}
Compared to source detection for scenario with complete nodes, the missing information from incomplete nodes may mistakenly be considered as valid data from normal nodes, leading to significant feature inaccuracies. Hence, distinguishing incomplete nodes is necessary. Additionally, the user heterogeneity and class imbalance problem in source detection hinder the effective fitting of models. Therefore, focusing on the nodes with greater information transmission capacity and differentiating between source and non-source sets becomes imperative.

\section{Method}
\if 0
This section provides a comprehensive overview of our proposed framework, i.e., Source Detection in Graphs with Incomplete Nodes via Positional Encoding and Attentive Fusion (GIN-SD). The overarching structure of GIN-SD is visualized in Fig. \ref{framework} and encompasses two fundamental modules: the Positional Embedding Module (PEM) and the Attentive Fusion Module (AFM).
\fi

In this section, we describe our proposed framework, i.e., Source Detection in Graphs with Incomplete Nodes via Positional Encoding and Attentive Fusion (GIN-SD). The framework consists of two primary components: the Positional Embedding Module (PEM) and the Attentive Fusion Module (AFM), as illustrated in Fig.~\ref{framework}.

Given the network snapshot $G'$ as input, PEM  embeds position-based user encodings, and then AFM  learns node representations through a self-attention mechanism. Finally, the loss is computed using a class balancing mechanism.

\subsection{Positional Embedding Module (PEM)}

% \subsubsection{Input of GIN-SD.}
% A batch size $b$ of network snapshots serves as the input for the entire model. For each sample $G'$, it includes the following information about user $i$: 1) user state, i.e., whether $v_i$ is in $G_+$ or $G_-$; 2) propagation information, i.e., the timestamp $t$ when the user received the corresponding message on the social network. Notably, $G'$ contains Incomplete nodes, which lack the aforementioned information.

% \subsubsection{Node Features Encoding.}
% During the node encoding phase, s
Several different perspectives of features, including user states, propagation information, and positional information, are embedded into the node feature vectors.

\subsubsection{User State Information $(\textbf{\textit{X}}_i^1)$}
When $\theta\%$ of users in the network receive the rumor information and are influenced by it, i.e., $|G_+| \geq \theta\% * n$, we obtain the network snapshot $G'$, in which the user states can be categorized into three sets: $G_+$, users influenced by the rumor; $G_-$, users not influenced or not receiving the rumor; and $\Psi$, users with lost information. Therefore, for user $v_i$, the state feature $\textbf{\textit{X}}_i^1$ can be determined by the following rules:
\begin{equation}
\textbf{\textit{X}}_i^1 = \left\{\begin{array}{ll}
+1, & v_i \in G_+ \\
-1, & v_i \in G_- \\
\ \ \ 0, & v_i \in \Psi.
\end{array}\right.
\end{equation}

\subsubsection{The Diffusion Information $(\textbf{\textit{X}}_i^2)$}
Social platforms like Facebook or Twitter include timestamps when users receive messages, which is a crucial factor in source detection. Therefore, for user $v_i$, in conjunction with the timestamp $t_i$, we define the diffusion information $\textbf{\textit{X}}_i^2$ as follows:
\begin{equation}
\textbf{\textit{X}}_i^2 = \left\{\begin{array}{ll}
\ \ t_i, & v_i \in G_+ \\
-1, & \text {otherwise}.
\end{array}\right.
\end{equation}

\subsubsection{The Positional Information $(\textbf{\textit{X}}_i^3)$}
Under the premise of node information loss, the inter-nodal positional relationships play a pivotal role in facilitating message propagation at the global level. To address this, leveraging the generalization of Laplacian positional encodings, we utilize it as the positional information embedded in the user feature $\textbf{\textit{X}}_i^3$ to distinguish incomplete nodes.

% Additionally, GCNs prioritize local learning and typically acquire structural node information with invariant to the node positions \cite{srinivasan2019equivalence}. Which is a critical reason why attention-based models like GATs struggle to achieve more competitive performance on graph datasets \cite{dwivedi2020generalization}.

In contrast to computing the Laplacian PEs for the entire network, we focus on calculating the Laplacian PEs for the infected subgraph $G_+'$. Given a network snapshot $G'$, if user $v_i$ did not receive the rumor or is not persuaded, the extraction process for the infected subgraph $G_+'$ is represented as:
\begin{equation}
\textbf{\textit{A}}_+ = \textbf{\textit{J}}_{i,n} \cdot \textbf{\textit{A}} \cdot \textbf{\textit{J}}_{i,n}^T,
\end{equation}
where $\textbf{\textit{A}}^{n \times n}$ is the adjacency matrix of the network snapshot $G'$, and $\textbf{\textit{A}}_+^{(n-1) \times (n-1)}$ is the adjacency matrix of the infected subgraph after removing user $v_i$, serving as the basis for subsequent removals. $\textbf{\textit{J}}_{i,n}$ denotes the $n$-dimensional identity matrix with its $i$-th row removed. For example, if the user with $id$-2 did not receive the rumor or not be persuaded,
\begin{equation}
\textbf{\textit{J}}_{2,n}^{(n-1) \times n}=\left(\begin{array}{ccccc}
1 & 0 & 0 & \cdots & 0 \\
0 & 0 & 1 & \cdots & 0 \\
\vdots & \vdots & \vdots & \ddots & \vdots \\
0 & 0 & 0 & \cdots & 1
\end{array}\right).
\end{equation}
It is essential to note that the users with unclear states in the $\Psi$ set are retained, meaning $G_+ \subset G_+'$, $\Psi \subset G_+'$, and $G_- \cap G_+' = \emptyset$.

After abstracting the infected subgraph, the symmetrically normalized Laplacian matrix is defined as:
\begin{equation}
\textit{\textbf{L}}_+^{sym} = \textbf{\textit{I}} - \textbf{\textit{D}}_+^{-1/2} \textbf{\textit{A}}_+ \textbf{\textit{D}}_+^{-1/2},
\end{equation}
where $\textbf{\textit{D}}_+$ is the degree matrix of infected subgraph. Subsequently, factorization is performed on matrix $\textit{\textbf{L}}_+^{sym}$:
\begin{equation}
\bigtriangleup_{\textit{\textbf{L}}_+^{sym}} = \textbf{\textit{$\Gamma$}}^T \textbf{\textit{$\lambda$}} \textbf{\textit{$\Gamma$}},
\end{equation}
$\Gamma$ and $\lambda$ represent the eigenvector and eigenvalue matrices of $\textit{\textbf{L}}_+^{sym}$, respectively. We select $k$ smallest non-trivial eigenvectors as $\Gamma_i$ for user $v_i$'s positional information $(k \ll n)$. In summary, the positional encoding $\textbf{\textit{X}}_i^3$ for user $v_i$ can be represented as:
\begin{equation}
\textbf{\textit{X}}_i^3 = \left\{\begin{array}{ll}
\Gamma_i, & v_i \in G_+'(V,E) \\
-1, & \text {otherwise}.
\end{array}\right.
\label{X_3}
\end{equation}

As the proposed framework follows a heuristic approach, the aforementioned user features can be further enriched. For instance, given a infected user $v_i$, in order to augment the discriminative capabilities, $\textbf{\textit{X}}_i^1$ may be defined as $\textbf{\textit{X}}_i^1 = (+1, -1)$. Furthermore, $\textbf{\textit{X}}_i^2$ can be extended to encompass both the timestamp of $v_i$ and the unique identifier ($id$) of the information propagator, denoted as $\textbf{\textit{X}}_i^2 = (t_i, id_i)$, where $id_i$ represents the $id$ of the individual responsible for disseminating the information to user $v_i$. It is imperative to emphasize that such an extensible feature engineering process fosters the exploration of richer information representation, thereby potentially enhancing the overall efficacy and robustness of the model in source detection tasks.

Finally, a concatenation procedure is employed to amalgamate the diverse user feature components, culminating in the derivation of the ultimate user embedding vector:
\begin{equation}
\textbf{\textit{X}}_i = \left[\|_{x=1}^3 \textbf{\textit{X}}_i^x\right].
\end{equation}

\subsection{Attentive Fusion Module (AFM)}
% \subsubsection{Self-Attention Mechanism.}
Considering the efficiency of information transmission varies among nodes, we focus on the nodes with greater information transmission capacity. Specifically, considering user $v_i$ and its neighbor $v_j$, the attention coefficient $e_{ij}$ at the $l$-th layer of model is formulated as:
\begin{equation}
e_{i j} = \Vec{a}\operatorname{LReLU}\left(\textbf{\textit{W}}^{(l)}\left( \textbf{\textit{X}}_{i}^{(l)}, \textbf{\textit{X}}_{j}^{(l)}\right)\right),
\label{eij}
\end{equation}
where $\textbf{\textit{X}}^{(l)} \in \mathbb{R}^{l_w \times n}$ is the feature representation of users and $\textbf{\textit{X}}^{(0)}$ = $\textbf{\textit{X}}$; $l_w$ signifies the number of elements in the node feature vector. $\textbf{\textit{W}}^{(l)} \in \mathbb{R}^{l_w' \times l_w}$ represents a trainable parameter matrix, $\operatorname{LReLU}(\cdot)$ is the activation function and $\Vec{a} \in \mathbb{R}^{2l_w'}$ is a weight vector.

Following the definition of $e_{ij}$, the weight of user $v_j$ concerning all neighbors of $v_i$ is computed as:
\begin{equation}
\alpha_{i j} = \frac{\exp \left(\vec{a}^{T}\operatorname{LReLU}\left(\textbf{\textit{W}}\left[ \textbf{\textit{X}}_{i} \| \textbf{\textit{X}}_{j}\right]\right)\right)}{\sum_{v_k \in N(v_i)} \exp \left(\vec{a}^{T}\operatorname{LReLU}\left(\textbf{\textit{W}}\left[ \textbf{\textit{X}}_{i} \| \textbf{\textit{X}}_{k}\right]\right)\right)},
\label{aij}
\end{equation}
where $N(v_i)$ denotes the neighbors of node $v_i$; $\|$ represents the vector concatenation operation; and $(\cdot)^T$ symbolizes the transposition. The final representation output for user $v_i$ is:
\begin{equation}
\textbf{\textit{X}}_{i}^{\prime} = \sum_{v_{j} \in N\left(v_{i}\right)} \alpha_{i j} \textbf{\textit{W}} \textbf{\textit{X}}_{j}.
\label{singlehead}
\end{equation}

In pursuit of augmenting the expressive power of the diffusion model and promoting the stability of the self-attention learning process, we deploy $K$ distinct and independent attention mechanisms, dedicated to capturing diverse aspects of information propagation. Subsequently, these $K$ mechanisms are concatenated to form a comprehensive and enriched representation:
\begin{equation}
   \textbf{\textit{X}}_{i}^{\prime\prime} = {\|}_{k = 1}^K \sigma\left(\textbf{\textit{X}}_{i}^{\prime k}\right).
   \label{multihead}
\end{equation}

To achieve dimension alignment, we perform a mean pooling operation on the individual attention channels at the ultimate layer of the model:
\begin{equation}
\textbf{\textit{X}}_{i}^{\prime\prime\prime} = \sigma\left(\frac{1}{K} \sum_{k = 1}^{K} \textbf{\textit{X}}_{i}^{\prime k}\right).
\end{equation}
This operation consolidates the diverse learned information from the attention mechanisms, harmonizing their representations and yielding a cohesive and coherent output for each node. Finally, the model yields an $(n \times 2)$-dimensional matrix, wherein each row's two elements undergo a $\operatorname{softmax}(\cdot)$ transformation:
\begin{equation}
S(\Vec{z})_i = \frac{e^{z_{i}}}{\sum_{j} e^{z_{j}}},\  \Vec{z} = \textbf{\textit{X}}_{i}^{\prime\prime\prime}.
\end{equation}

% Note that we do not aim at designing a new attention mechanism, but instead propose a novel module for assigning attention coefficients to nodes, including incomplete ones, based on their information transmission capacity as our main contributions.

It is important to emphasize that our focus is not on devising a novel attention mechanism, but rather on introducing an innovative attentive fusion module. This module aims to allocate attention coefficients to nodes dynamically, encompassing those that are incomplete, contingent on their information transmission capacity. This constitutes our primary contributions in this context.

\subsection{Loss Function and Training}
% In the context of social platforms, large-scale rumor propagation predominantly originates from a few sources, whereas the social network itself exhibits a vast scale of millions to billions of users, leading to a substantial disparity between source and non-source set (i.e., $n \gg |s|$) thereby giving rise to the class imbalance problem. This issue often prompts models to excessively prioritize the larger set while neglecting the smaller set, resulting the prediction bias.

To rectify the class imbalance problem and ensure equitable attention across all sets, we propose a class-balancing mechanism. For the source set $s$ and the non-source set $V-s$, we introduce a fixed constant $\xi$:
\begin{equation}
\xi = \frac{|s|}{n-|s|},
\end{equation}
where $n$ and $|s|$ represent the number of elements in sets $V$ and $s$ respectively. The constant $\xi$ equalizes the weights of all samples and align their mathematical expectations to 1, thereby promoting unbiased and comprehensive learning.

Through integrating this class-balancing mechanism into our model, we construct a novel loss function that is formulated as follows:
\begin{equation}
Loss = \sum_{v_{i} \in s} L_{i}+\xi \sum_{v_{j} \in(V-s)} L_{j}+\lambda\|w\|_{2},
\label{loss}
\end{equation}
where $L$ represents the cross-entropy loss; for sample $x$ and its label $y$, $L(x, y)$ = $-\log (x) \times y$. The last term in Eq. (\ref{loss}) denotes the $L_2$ regularization.

% Furthermore, the class-balancing mechanism above exhibits the potential to be extended and adapted for handling class-imbalanced classification problems across diverse domains and applications, providing a versatile solution to address the challenge of imbalanced data distributions.

The integrated GIN-SD, incorporating PEM and AFM, focuses on distinguishing incomplete nodes while prioritizing nodes with higher information transmission capacity. Additionally, the class balancing mechanism further ensures differential treatment of source/non-source sets. This synergy enables effective information extraction and efficient source detection.

\section{Experiments}

\subsection{Experimental Setting}

\subsubsection{Implementation}
Given the independent nature of each user's social behavior and the short-term property of rumors, we randomly select 5$\%$ of the users as sources to construct incomplete graph about rumor propagation. Then, we employ the heterogeneous Independent Cascade (IC) model to simulate rumor dissemination, where each user's forwarding probability $p$ is drawn from a uniform distribution $U(0.1, 0.5)$. The propagation is halted when 30$\%$ of the users are influenced by the rumor, and the network snapshot is obtained with proportion of $\delta$ incomplete nodes. The training and testing set have a sample ratio of 8 : 2 and the learning rate is set to $10^{-3}$. The number of attention layer equals to 3.
For small-scale networks $(G_1$-$G_2)$, the number of attention heads is set to 4, and the number of neurons in the hidden layer is 800. For medium-scale networks $(G_3$-$G_7)$, the corresponding numbers are set to 2 and 500 respectively for the consideration computational constraints. As to the large-scale network $(G_8)$, the number of attention heads is assigned to 1, and the number of neurons in the hidden layer equals to 500.
All experiments are conducted on a workstation with a single NVIDIA RTX 3090Ti GPU.

\subsubsection{Datasets}
Eight real-world datasets of different scales are utilized to evaluate the performance of each method, including Football \cite{girvan2002community}, Jazz \cite{gleiser2003community}, Facebook \cite{leskovec2012learning}, Twitch-ES \cite{rozemberczki2021multi}, LastFM \cite{rozemberczki2020characteristic}, Enron \cite{klimt2004enron}, Github \cite{rozemberczki2021multi} and DBLP \cite{yang2012defining}. The specific characteristics are presented in Table \ref{table_network}.

\subsubsection{Evaluation Metrics}
The widely used Accuracy (Acc) and F-Score \cite{wang2017multiple} are selected as the fundamental evaluation metrics to assess the efficacy of the methods. Acc quantifies the proportion of correctly classified samples among the entire sample population, while F-Score comprises two components: $Precision$ and $Recall$. $Precision$ quantifies the proportion of true sources within $\hat{s}$, denoted as $|\hat{s} \cap s|\ /\ |\hat{s}|$, while $Recall$ gauges the proportion of detected sources in $s$, represented as $|\hat{s} \cap s|\ /\ |s|$. These metrics provide a comprehensive and rigorous assessment of the methods' performance in capturing the veracity and completeness of source detection results.

\begin{table}[!h]
	\renewcommand{\arraystretch}{1.0}
	\centering
	\begin{tabular}{lllll}
		\toprule	
		\multicolumn{2}{l}{Network} & $|V|$ & $|E|$ & $\langle k \rangle$ \\
		\midrule
		$G_1$ & Football & 115 & 613 & 10.66 \\
		$G_2$ & Jazz & 198 & 2742 & 27.70 \\
		$G_3$ & Facebook & 4039 & 88234 & 43.69 \\
		$G_4$ & Twitch-ES & 4648 & 59382 & 25.55 \\
		$G_5$ & LastFM & 7624 & 27806 & 7.29 \\
	    $G_6$ & Enron & 36692 & 183831 & 10.02 \\
        $G_7$ & Github & 37700 & 289003 & 15.33 \\
        $G_8$ & DBLP & 317080 & 1049866 & 6.62 \\
		\bottomrule
	\end{tabular}
 \caption{Characteristics of the considered datasets.}%标注该表格，用于在文章内引用
 \label{table_network}
\end{table}

\begin{table*}[t]
\renewcommand{\arraystretch}{1.0} %高度
\setlength\tabcolsep{4pt} %宽度
\centering
% \resizebox{\textwidth}{!}{%
\begin{tabular}{lcccccccccccc}
% \toprule
\hline
                 & \multicolumn{2}{c}{\textbf{Football}}  & \multicolumn{2}{c}{\textbf{Jazz}}      & \multicolumn{2}{c}{\textbf{Facebook}}  & \multicolumn{2}{c}{\textbf{Twitch-ES}}    & \multicolumn{2}{c}{\textbf{Github}}    & \multicolumn{2}{c}{\textbf{DBLP}}      \\ \cmidrule(r){2-3} \cmidrule(r){4-5} \cmidrule(r){6-7} \cmidrule(r){8-9}  \cmidrule(r){10-11} \cmidrule(r){12-13}
\textbf{Methods} & \textbf{Acc} & \textbf{F-Score} & \textbf{Acc} & \textbf{F-Score} & \textbf{Acc} & \textbf{F-Score} & \textbf{Acc} & \textbf{F-Score} & \textbf{Acc} & \textbf{F-Score} & \textbf{Acc} & \textbf{F-Score} \\ \hline
LPSI            &          0.812        &         0.323            &         0.794         &           0.302          &         0.811         &           0.014          &         0.795         &           0.008          &          0.783        &           0.002          &         0.755         &          0           \\
EPA            &          0.783        &           0.303          &         0.806         &          0.295           &          0.798        &           0.010          &          0.783        &          0.002           &         0.792         &           0.001          &          0.763        &          0           \\
GCNSI             &         0.831         &          0.284           &         0.829         &          0.271           &          0.835        &           0.004          &          0.820        &          0.003           &         0.811         &           0.001          &         0.807         &          0           \\
SIGN             &         0.809         &          0.513           &         0.794         &          0.495           &          0.819        &           0.452          &          0.790        &          0.443           &         0.775         &           0.373          &         0.768         &          0.248           \\
GCSSI             &         0.779         &          0.495           &         0.786         &          0.447           &          0.807        &           0.423          &          0.797        &          0.427           &         0.783         &           0.386          &         0.771         &          0.265           \\
ResGCN             &         0.824         &          0.502           &         0.795         &          0.475           &          0.816        &           0.440          &          0.823        &          0.429           &         0.790         &           0.379          &         0.785         &          0.251           \\
IVGD             &         0.897         &          0.729           &         0.904         &          0.684           &          0.882        &           0.661          &          0.837        &          0.625           &         0.819         &           0.580          &         0.804         &          0.533           \\
SL-VAE             &         0.887         &          0.716           &         0.846         &          0.672           &          0.865        &           0.651          &          0.827        &          0.618           &         0.803         &           0.542          &         0.810         &          0.516           \\ \hline
GIN-SD     &         \textbf{0.956}         &          \textbf{0.839}           &         \textbf{0.934}         &           \textbf{0.715}          &         \textbf{0.968}         &           \textbf{0.761}          &         \textbf{0.970}         &           \textbf{0.764}          &          \textbf{0.912}        &         \textbf{0.694}            &         \textbf{0.895}         &          \textbf{0.690}          \\ \hline
\end{tabular}%
% }
\caption{Source detection performance in graphs with 10$\%$ incomplete nodes. The best results are highlighted in bold.}
\label{overallperformance1}
\end{table*}

\begin{table*}[t]
\renewcommand{\arraystretch}{1.0} %高度
\setlength\tabcolsep{4pt} %宽度
\centering
% \resizebox{\textwidth}{!}{%
\begin{tabular}{lcccccccccccc}
% \toprule
\hline
                 & \multicolumn{2}{c}{\textbf{Football}}  & \multicolumn{2}{c}{\textbf{Jazz}}      & \multicolumn{2}{c}{\textbf{LastFM}}  & \multicolumn{2}{c}{\textbf{Enron}}    & \multicolumn{2}{c}{\textbf{Github}}    & \multicolumn{2}{c}{\textbf{DBLP}}      \\ \cmidrule(r){2-3} \cmidrule(r){4-5} \cmidrule(r){6-7} \cmidrule(r){8-9}  \cmidrule(r){10-11} \cmidrule(r){12-13}
\textbf{Methods} & \textbf{Acc} & \textbf{F-Score} & \textbf{Acc} & \textbf{F-Score} & \textbf{Acc} & \textbf{F-Score} & \textbf{Acc} & \textbf{F-Score} & \textbf{Acc} & \textbf{F-Score} & \textbf{Acc} & \textbf{F-Score} \\ \hline
LPSI            &          0.798        &         0.206            &         0.775         &           0.220          &         0.792         &           0.006          &         0.773         &           0.001          &          0.751        &           0          &         0.733         &          0           \\
EPA            &          0.771        &           0.195          &         0.759         &          0.216           &          0.768        &           0.007          &          0.752        &          0           &         0.738         &           0          &          0.718        &          0           \\
GCNSI             &         0.782         &          0.157           &         0.779         &          0.176           &          0.780        &           0.001          &          0.761        &          0           &         0.750         &           0          &         0.746         &          0           \\
SIGN             &         0.795         &          0.282           &         0.784         &          0.253           &          0.802        &           0.231          &          0.773        &          0.217           &         0.762         &           0.205          &         0.763         &          0.184           \\
GCSSI             &         0.772         &          0.259           &         0.785         &          0.236           &          0.795        &           0.210          &          0.781        &          0.201           &         0.778         &           0.194          &         0.764         &          0.150           \\
ResGCN             &         0.816         &          0.264           &         0.782         &          0.241           &          0.804        &           0.227          &          0.814        &          0.212           &         0.783         &           0.215          &         0.781         &          0.164           \\
IVGD             &         0.872         &          0.506           &         0.859         &          0.496           &          0.867        &           0.509          &          0.820        &          0.426           &         0.809         &           0.424          &         0.780         &          0.413           \\
SL-VAE             &         0.874         &          0.492           &         0.839         &          0.501           &          0.846        &           0.516          &          0.813        &          0.447           &         0.791         &           0.415          &         0.785         &          0.398           \\ \hline
GIN-SD     &         \textbf{0.897}         &          \textbf{0.721}           &         \textbf{0.904}         &           \textbf{0.635}          &         \textbf{0.914}         &           \textbf{0.657}          &         \textbf{0.921}         &           \textbf{0.694}          &          \textbf{0.854}        &         \textbf{0.605}            &         \textbf{0.846}         &          \textbf{0.613}          \\ \hline
\end{tabular}%
% }
\caption{Source detection performance in graphs with 20$\%$ incomplete nodes. The best results are highlighted in bold.}
\vspace{-4mm}
\label{overallperformance2}
\end{table*}

% and F-Score is calculated as:
% \begin{equation}
%     F\verb|-|Score = (1+\beta^2)*\frac{Precision*Recall}{\beta^{2}*Precision+Recall},
% \end{equation}
% the parameter $\beta$ is set to 1 (i.e., F1-Score).

\subsubsection{Baselines}
Eight recently proposed  representative source detection methods are considered as baselines, including LPSI \cite{wang2017multiple} and EPA \cite{ali2019epa} based on source centrality theory; GCNSI \cite{dong2019multiple}, SIGN \cite{li2021propagation}, GCSSI \cite{dong2022wavefront} and ResGCN \cite{shah2020finding} that consider user states; IVGD \cite{wang2022invertible} and SL-VAE \cite{ling2022source} which incorporate user and propagation information.

% Eight representative source detection methods proposed recent years are selected as baselines: LPSI \cite{wang2017multiple} adopts an iterative label value propagation approach to identify locally prominent nodes as sources. EPA \cite{ali2019epa} determines the source node by iteratively computing the infection time for each node and selecting the node with the longest infection time. GCNSI \cite{dong2019multiple} leverages the output of LPSI as node features and employs an $n$-classification scheme to determine the final source node. SIGN \cite{li2021propagation} learns node representations based on node states through label propagation. GCSSI \cite{dong2022wavefront} concentrates on the last infected users, i.e., the wavefront, and proposes a sequence-to-sequence model for source identification. IVGD \cite{wang2022invertible} endeavors to locate sources by reconstructing the propagation process through reverse graph diffusion. SL-VAE \cite{ling2022source} exploits a stochastic model with forward diffusion to approximate the source distribution effectively. Bi-GCN \cite{bian2020rumor} enhances the impact of source post at each layer by capitalizing on the top-down and bottom-up structures in the model.

\subsection{Comparison with State-of-the-art Methods}
To validate the effectiveness of GIN-SD, we conduct comprehensive comparisons with benchmark methods on eight datasets $(G_1$-$G_8)$ for two scenarios with $\delta$ being equivalent to 0.1 and 0.2. The results are summarized as in Table \ref{overallperformance1} and Table \ref{overallperformance2} respectively. Through the experimental analysis, we have derived several key observations:

Firstly, all methods exhibit commendable Acc performance, whereas the F-Score appears relatively lower. This disparity stems from the class imbalance issue, where non-source samples significantly outnumber the source samples. In other words, the larger the difference between Acc and F-Score, the more the model is affected by the class imbalance problem. Notably, three benchmark methods (LPSI, EPA, and GCNSI) exhibit relatively typical performance characteristics. Furthermore, models that incorporate the learning mechanism of information diffusion processes outperform their counterparts, as evidenced by the significantly superior performance of the latter three methods compared to the initial six. Additionally, the influence of user information loss is evident, as all benchmark methods manifest a substantial decline in performance compared to their optimal results. This decline stems from the challenge posed by incomplete nodes, hindering the simulations' convergence and yielding errors in the model's predictions. In conclusion, among all methods, GIN-SD emerges as the optimal performer. Notably, in contrast to models that overlook the propagation process, GIN-SD exhibits an average improvement of $32\%$, and the enhancement ranges from $5\%$ to $18\%$ based on the models consider the propagation process. This substantial improvement is attributed to the salient enhancements introduced by GIN-SD, including: 1) leveraging positional information to distinguish incomplete nodes, 2) employing attention mechanism to enable the model's targeted focus on distinct nodes, and 3) introducing a class-balancing mechanism to tackle the class imbalance problem.

% In other words, the larger the difference between Acc and F-Score, the more the model is affected by the class imbalance problem. Notably, the benchmark methods LPSI, EPA, and GCNSI exhibit relatively typical performance characteristics. Furthermore, models that incorporate the learning of information diffusion processes outperform their counterparts, as evidenced by the significantly superior performance of the latter three methods compared to the initial five. This distinction can be attributed to the stochastic nature of information propagation, which necessitates a more comprehensive consideration beyond mere topological structure and user state information to effectively capture the specific propagation characteristics. Additionally, the influence of user information loss is evident, as all benchmark methods manifest a substantial decline in performance relative to their optimal results. This decline stems from the challenge posed by incomplete nodes, hindering the simulations' convergence and yielding errors in the model's predictions.

\subsection{Performance on Early Rumor Sources Detection}
\begin{figure}[htbp]
  \centering
  \includegraphics[width=\linewidth]{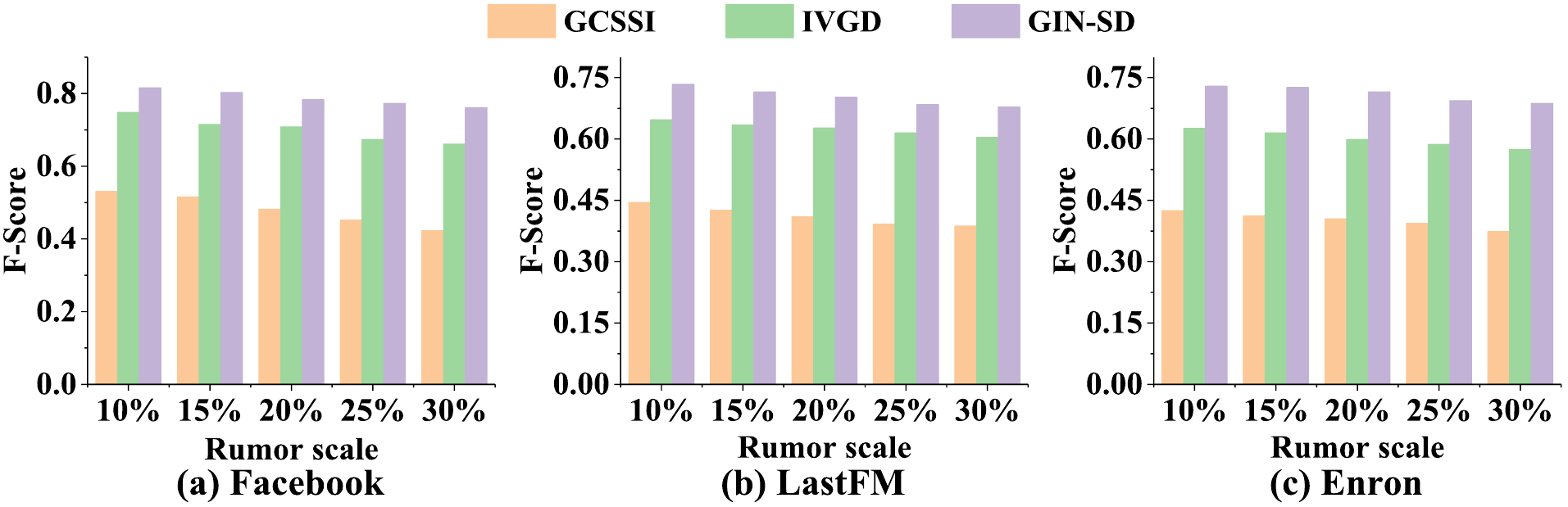}
  \caption{The performance of different methods in early rumor sources detection.}
  \vspace{-1mm}
  \label{early_detection}
  \vspace{-5mm}
\end{figure}

Due to the amplified and persistent harm incurred by the rumors propagation in society, it is of vital significance to identify the sources at the early stages of rumor dissemination to curtail further spreading. To evaluate the efficacy of distinct methodologies in the context of early rumor sources detection, we initial the source detection procedure when the rumor's influence extends to 10$\%$ to 30$\%$ of users, with an increment of 5$\%$. The incomplete node ratio $\delta$ is set to 0.1. The results are presented in Fig. \ref{early_detection}.
% devise an experimental setup with varying rumor propagation scales. Specifically, we initial the source detection procedure when the rumor's influence extends to 10$\%$ to 30$\%$ of users, with an increment of 5$\%$. And $\delta$ is set to 0.1. The results are visually presented in Fig. \ref{early_detection}.

% To empirically evaluate the efficacy of distinct methodologies in the context of early rumor sources detection, we devise an experimental setup with varying rumor propagation scales. Specifically, we initial the source detection procedure when the rumor's influence extends to 10$\%$ to 30$\%$ of users, with an increment of 5$\%$. And 10$\%$ of users experiencing information loss. The results of the source detection task are visually presented in Fig. \ref{early_detection}.

The findings reveal a discernible trend: as the scale of rumors expands, all methods exhibit a decrease in source detection precision. This underscores the imperative of timely source detection during the early phases of rumor propagation, considering both the potential societal ramifications and the inherent challenges in identifying sources. Moreover, across diverse scenarios, GIN-SD consistently attains the highest level of source detection precision, serving as empirical evidence supporting the efficacy and rationale of GIN-SD's class-balancing mechanism, as well as its incorporation of PEM and AFM modules.

\begin{table}[b]
\vspace{-1mm}
\renewcommand{\arraystretch}{1.0} %高度
\setlength\tabcolsep{2pt} %宽度
\centering
% \resizebox{.5\textwidth}{!}{%
\begin{tabular}{lcccccc}
% \toprule
\hline
                 & \multicolumn{2}{c}{\textbf{Facebook}}  & \multicolumn{2}{c}{\textbf{LastFM}} & \multicolumn{2}{c}{\textbf{Github}}      \\ \cmidrule(r){2-3} \cmidrule(r){4-5} \cmidrule(r){6-7}
\textbf{Methods} & \textbf{Acc} & \textbf{F-Score} & \textbf{Acc} & \textbf{F-Score} & \textbf{Acc} & \textbf{F-Score}\\ \hline
w/o P            &          0.823        &         0.413            &         0.815         &           0.404           &         0.798         &           0.339\\
w/ P'            &          0.426        &         -            &         0.415         &           -           &         0.397         &           -\\ \hline
w/o A            &          0.942        &           0.743          &         0.923         &          0.705           &         0.908         &           0.651\\
w/ AS            &          0.801        &           0.223          &         0.782         &          0.210           &         0.769         &           0.179\\
w/ AL            &          0.907        &           0.716          &         0.891         &          0.683           &         0.854         &           0.617\\ \hline
w/o B             &         0.825         &          0.223           &         0.817         &          0.205            &         0.809         &           0.198\\ \hline
GIN-SD     &         \textbf{0.968}         &          \textbf{0.761}           &         \textbf{0.950}         &           \textbf{0.726}    &         \textbf{0.912}         &           \textbf{0.694}\\ \hline
\end{tabular}%
% }
\vspace{-1mm}
\caption{The performance of different variants for GIN-SD.}
\label{ablation}
\vspace{-3mm}
\end{table}

\subsection{Impact of Incomplete Ratio}
\begin{figure}[t]
  \vspace{-1mm}
  \centering
  \includegraphics[width=\linewidth]{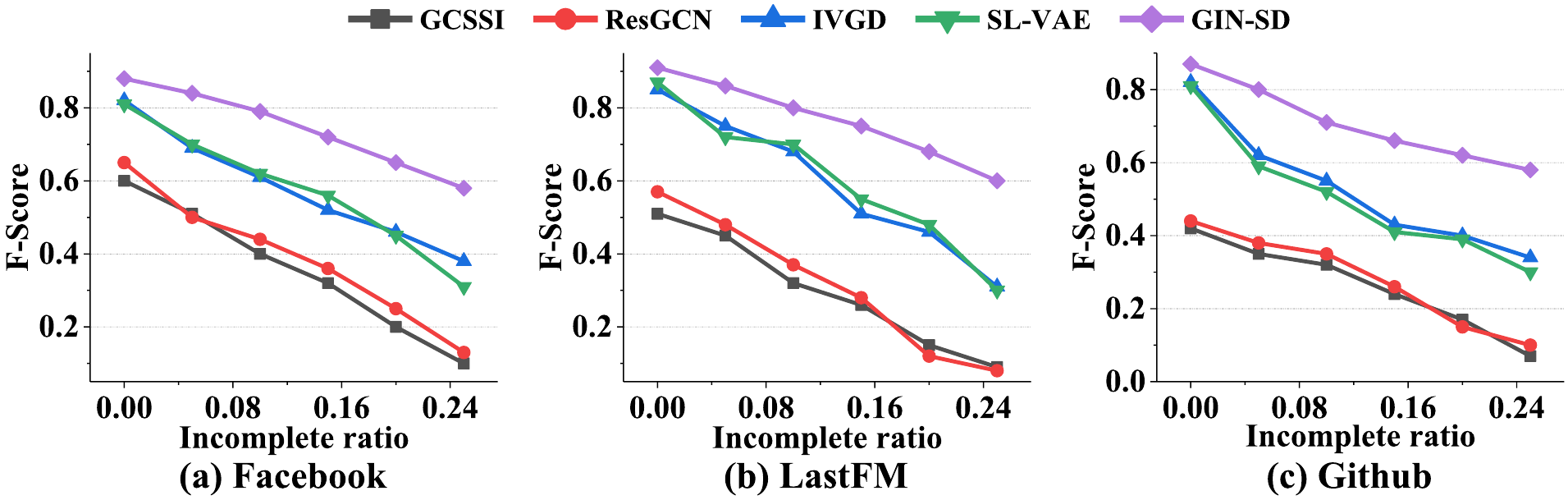}
  \vspace{-5mm}
  \caption{The impact of varying degrees of user information loss on source detection accuracy.}
  \label{incomplete_ratio}
  \vspace{-4mm}
\end{figure}

To further explore the effects of user information loss on source detection, we systematically vary the incomplete node ratio, i.e., $\delta$ in the range of 0 to 0.25 with a step size 0.05. The experimental results are depicted as in Fig. \ref{incomplete_ratio}.

The results reveal a pronounced degradation in source detection accuracy for all methods as incomplete nodes intensifies. This deterioration is primarily attributed to the perturbation caused by user information loss, impeding the effective convergence of the models. Furthermore, the accuracy of the GCSSI method, which does not consider the propagation process, is notably lower than other considered methods. In contrast, GIN-SD exhibits a remarkable superiority which amplifies with the increase of $\delta$, thus substantiating its high applicability and resilience.

% Furthermore, the accuracy of the GCSSI method, which does not consider the propagation process, is notably lower compared to other considered methods.

\subsection{Ablation Study and Analysis}
To validate the necessity of each module in GIN-SD, we conduct ablation studies targeting its components and summarize the results  in Table \ref{ablation}. The variants are designed as:

% The key modules of GIN-SD include: enhanced global learning through position encoding, reinforced local learning via self-attention mechanism, and the class-balancing mechanism. Therefore, the variants are designed as:

\begin{itemize}
    \item \textbf{GIN-SD w/o P} utilizes a zero vector to replace the user's positional encodings, i.e., $\textbf{\textit{X}}_i^3 = [0]^{n \times k}$ in Eq. (\ref{X_3}).
    \item \textbf{GIN-SD w/ P'} calculates the proportion of correctly identified nodes among those that are sources and have missing information, i.e., $\left(\hat{s} \cap \Psi\right)/\left(s \cap \Psi\right)$.
    \item \textbf{GIN-SD w/o A} removes the attention mechanism, i.e., $\textbf{\textit{X}}_i^{\prime} = \sum_{v_{j} \in N\left(v_{i}\right)} \textbf{\textit{W}} \textbf{\textit{X}}_{j}$ in Eq. (\ref{singlehead}).
    \item \textbf{GIN-SD w/ AS} assigns higher attention weights to nodes with smaller degrees, i.e., $\alpha_{ij} \propto 1 / |N(v_j)|$ in Eq. (\ref{aij}).
    \item \textbf{GIN-SD w/ AL} assigns higher attention weights to nodes with larger degrees, i.e., $\alpha_{ij} \propto |N(v_j)|$ in Eq. (\ref{aij}).
    \item \textbf{GIN-SD w/o B} removes the class-balancing mechanism, i.e., $\xi = 1$ in Eq. (\ref{loss}).
\end{itemize}

\subsubsection{Positional Embedding}
We validate the importance of positional embedding through evaluating the variants GIN-SD w/o P and GIN-SD w/ P'; according to the experimental outcomes, the impact of incomplete nodes on GIN-SD w/o P is prominently pronounced, resulting in discernible deviations in the model's accuracy. Moreover, the performance of GIN-SD w/ P' in accurately discerning incomplete source nodes to a certain degree underscores the efficacy of incorporating positional information.
\vspace{-1mm}

\subsubsection{Attentive Fusion} We investigate the significance of attentive fusion by comparing GIN-SD with GIN-SD w/o A, GIN-SD w/ AS and GIN-SD w/ AL; based on their performance, GIN-SD w/ AS performs the worst due to an excessive focus on nodes with small degrees and limited information transmission capabilities. While GIN-SD w/ AL exhibits a higher proficiency, however, the presence of bridge nodes with strong information transmission capacity but not necessarily high degrees \cite{beers2023followback} limits its performance. Despite the exceptional performance of GIN-SD w/o A within the variant range, its uniform attention coefficients prevent it from reaching the superior capabilities demonstrated by the baseline GIN-SD framework.
\vspace{-1mm}

\subsubsection{Class-balancing Mechanism} The importance of class-balancing mechanism is validated through the relatively inferior performance of GIN-SD w/o B amongst the entire array of variants, which underscores the critical role played by the class balance mechanism in the source detection task.
\vspace{-2mm}

% The ablation studies confirm the necessity of each module in GIN-SD. More specifically, the impact of incomplete nodes on GIN-SD w/o P is prominently pronounced, resulting in discernible deviations in the model's accuracy. Moreover, the performance of GIN-SD w/ P' in accurately discerning incomplete source nodes to a certain degree underscores the efficacy of incorporating positional information. Turning to the attention variants, GIN-SD w/ AS performs the worst due to an excessive focus on nodes with small degrees. While GIN-SD w/ AL exhibits a higher proficiency, however, the presence of bridge nodes with strong information transmission capacity but not necessarily high degrees \cite{beers2023followback} limits its performance. Despite the exceptional performance of GIN-SD w/o A within the variant range, its uniform attention coefficients prevent it from reaching the superior capabilities demonstrated by the baseline GIN-SD framework. Lastly, amongst the entire array of variants, the relatively inferior performance of GIN-SD w/o B underscores the critical role played by the class balance mechanism in the source detection task.

% Among them, GIN-SD w/o B exhibits the poorest performance, highlighting the importance of the class-balancing mechanism in the source detection task. Subsequently, GIN-SD w/o P is significantly affected by incomplete nodes, resulting in larger model errors. While GIN-SD w/o A performs the best among the variants, it still falls short of the outstanding performance exhibited by GIN-SD.

\section{Conclusion}
This paper poses a new challenge for rumor source detection in graphs with incomplete nodes and has proposed a novel framework, GIN-SD, to tackle this problem. The key idea involves distinguishing incomplete nodes by leveraging position-based encoding of user features, followed by adaptive allocation of attention coefficients using a self-attention mechanism based on information transmission capacity. Additionally, a class balancing mechanism is devised to address prediction bias in the model. Extensive experimental results validate the effectiveness and superiority of our solution. We hope that this work, which introduces a new dimension to the field, will inspire further researches into robust deep learning models for source detection.

\section{Acknowledgments}
This work was supported by the National Key R\&D Program (no. 2022YFE0112300); the National Natural Science Foundation for Distinguished Young Scholars (no. 62025602); the National Natural Science Foundation of China (nos. U22B2036, 62261136549, 11931015 and 62073263); the Fok Ying-Tong Education Foundation, China (no. 171105); the Innovation Foundation for Doctor Dissertation of Northwestern Polytechnical University (no. CX2023068); and the Tencent Foundation and XPLORER PRIZE.

\bibliography{aaai24}

\begin{thebibliography}{37}
\providecommand{\natexlab}[1]{#1}

\bibitem[{Ali et~al.(2019)Ali, Anwar, Rastogi, and Rizvi}]{ali2019epa}
Ali, S.~S.; Anwar, T.; Rastogi, A.; and Rizvi, S. A.~M. 2019.
\newblock EPA: Exoneration and prominence based age for infection source identification.
\newblock In \emph{Proceedings of the 28th ACM International Conference on Information and Knowledge Management}, 891--900.

\bibitem[{Barth{\'e}lemy et~al.(2004)Barth{\'e}lemy, Barrat, Pastor-Satorras, and Vespignani}]{barthelemy2004velocity}
Barth{\'e}lemy, M.; Barrat, A.; Pastor-Satorras, R.; and Vespignani, A. 2004.
\newblock Velocity and hierarchical spread of epidemic outbreaks in scale-free networks.
\newblock \emph{Physical Review Letters}, 92(17): 178701.

\bibitem[{Beers et~al.(2023)Beers, Schafer, Kennedy, Wack, Spiro, and Starbird}]{beers2023followback}
Beers, A.; Schafer, J.~S.; Kennedy, I.; Wack, M.; Spiro, E.~S.; and Starbird, K. 2023.
\newblock Followback Clusters, Satellite Audiences, and Bridge Nodes: Coengagement Networks for the 2020 US Election.
\newblock In \emph{Proceedings of the International AAAI Conference on Web and Social Media}, volume~17, 59--71.

\bibitem[{Belkin and Niyogi(2003)}]{belkin2003laplacian}
Belkin, M.; and Niyogi, P. 2003.
\newblock Laplacian eigenmaps for dimensionality reduction and data representation.
\newblock \emph{Neural Computation}, 15(6): 1373--1396.

\bibitem[{Bian et~al.(2020)Bian, Xiao, Xu, Zhao, Huang, Rong, and Huang}]{bian2020rumor}
Bian, T.; Xiao, X.; Xu, T.; Zhao, P.; Huang, W.; Rong, Y.; and Huang, J. 2020.
\newblock Rumor detection on social media with bi-directional graph convolutional networks.
\newblock In \emph{Proceedings of the AAAI Conference on Artificial Intelligence}, volume~34, 549--556.

\bibitem[{Cheng et~al.(2022)Cheng, Li, Han, Luo, Ma, and Zhu}]{cheng2022path}
Cheng, L.; Li, X.; Han, Z.; Luo, T.; Ma, L.; and Zhu, P. 2022.
\newblock Path-based multi-sources localization in multiplex networks.
\newblock \emph{Chaos, Solitons \& Fractals}, 159: 112139.

\bibitem[{Dong et~al.(2022)Dong, Zheng, Li, Li, Zheng, and Zhou}]{dong2022wavefront}
Dong, M.; Zheng, B.; Li, G.; Li, C.; Zheng, K.; and Zhou, X. 2022.
\newblock Wavefront-Based Multiple Rumor Sources Identification by Multi-Task Learning.
\newblock \emph{IEEE Transactions on Emerging Topics in Computational Intelligence}, 6(5): 1068--1078.

\bibitem[{Dong et~al.(2019)Dong, Zheng, Quoc Viet~Hung, Su, and Li}]{dong2019multiple}
Dong, M.; Zheng, B.; Quoc Viet~Hung, N.; Su, H.; and Li, G. 2019.
\newblock Multiple rumor source detection with graph convolutional networks.
\newblock In \emph{Proceedings of the 28th ACM International Conference on Information and Knowledge Management}, 569--578.

\bibitem[{Du et~al.(2017)Du, Jiang, Chen, Ren, and Poor}]{du2017community}
Du, J.; Jiang, C.; Chen, K.-C.; Ren, Y.; and Poor, H.~V. 2017.
\newblock Community-structured evolutionary game for privacy protection in social networks.
\newblock \emph{IEEE Transactions on Information Forensics and Security}, 13(3): 574--589.

\bibitem[{Dwivedi et~al.(2020)Dwivedi, Joshi, Luu, Laurent, Bengio, and Bresson}]{dwivedi2020benchmarking}
Dwivedi, V.~P.; Joshi, C.~K.; Luu, A.~T.; Laurent, T.; Bengio, Y.; and Bresson, X. 2020.
\newblock Benchmarking graph neural networks.
\newblock \emph{arXiv preprint arXiv:2003.00982}.

\bibitem[{Gao et~al.(2022)Gao, Zhu, Zhang, Wang, and Li}]{gao2022novel}
Gao, C.; Zhu, J.; Zhang, F.; Wang, Z.; and Li, X. 2022.
\newblock A novel representation learning for dynamic graphs based on graph convolutional networks.
\newblock \emph{IEEE Transactions on Cybernetics}, 53(6): 3599--3612.

\bibitem[{Girvan and Newman(2002)}]{girvan2002community}
Girvan, M.; and Newman, M.~E. 2002.
\newblock Community structure in social and biological networks.
\newblock \emph{Proceedings of the National Academy of Sciences}, 99(12): 7821--7826.

\bibitem[{Gleiser and Danon(2003)}]{gleiser2003community}
Gleiser, P.~M.; and Danon, L. 2003.
\newblock Community structure in jazz.
\newblock \emph{Advances in Complex Systems}, 6(04): 565--573.

\bibitem[{Han et~al.(2021)Han, Xiao, Wu, Guo, Xu, and Wang}]{han2021transformer}
Han, K.; Xiao, A.; Wu, E.; Guo, J.; Xu, C.; and Wang, Y. 2021.
\newblock Transformer in transformer.
\newblock \emph{Advances in Neural Information Processing Systems}, 34: 15908--15919.

\bibitem[{Kipf and Welling(2017)}]{kipf2016semi}
Kipf, T.~N.; and Welling, M. 2017.
\newblock Semi-supervised classification with graph convolutional networks.
\newblock In \emph{5th International Conference on Learning Representations, ICLR}.

\bibitem[{Klimt and Yang(2004)}]{klimt2004enron}
Klimt, B.; and Yang, Y. 2004.
\newblock The enron corpus: A new dataset for email classification research.
\newblock In \emph{Machine Learning: ECML 2004: 15th European Conference on Machine Learning, Pisa, Italy, September 20-24, 2004. Proceedings 15}, 217--226. Springer.

\bibitem[{Leskovec and Mcauley(2012)}]{leskovec2012learning}
Leskovec, J.; and Mcauley, J. 2012.
\newblock Learning to discover social circles in ego networks.
\newblock \emph{Advances in Neural Information Processing Systems}, 25.

\bibitem[{Li et~al.(2021)Li, Zhou, Jiang, and Huang}]{li2021propagation}
Li, L.; Zhou, J.; Jiang, Y.; and Huang, B. 2021.
\newblock Propagation source identification of infectious diseases with graph convolutional networks.
\newblock \emph{Journal of Biomedical Informatics}, 116: 103720.

\bibitem[{Ling et~al.(2022)Ling, Jiang, Wang, and Liang}]{ling2022source}
Ling, C.; Jiang, J.; Wang, J.; and Liang, Z. 2022.
\newblock Source localization of graph diffusion via variational autoencoders for graph inverse problems.
\newblock In \emph{Proceedings of the 28th ACM SIGKDD Conference on Knowledge Discovery and Data Mining}, 1010--1020.

\bibitem[{Parshani, Carmi, and Havlin(2010)}]{parshani2010epidemic}
Parshani, R.; Carmi, S.; and Havlin, S. 2010.
\newblock Epidemic threshold for the susceptible-infectious-susceptible model on random networks.
\newblock \emph{Physical Review Letters}, 104(25): 258701.

\bibitem[{Pinto, Thiran, and Vetterli(2012)}]{pinto2012locating}
Pinto, P.~C.; Thiran, P.; and Vetterli, M. 2012.
\newblock Locating the source of diffusion in large-scale networks.
\newblock \emph{Physical Review Letters}, 109(6): 068702.

\bibitem[{Prakash, Vreeken, and Faloutsos(2012)}]{prakash2012spotting}
Prakash, B.~A.; Vreeken, J.; and Faloutsos, C. 2012.
\newblock Spotting culprits in epidemics: How many and which ones?
\newblock In \emph{2012 IEEE 12th International Conference on Data Mining}, 11--20. IEEE.

\bibitem[{Rozemberczki, Allen, and Sarkar(2021)}]{rozemberczki2021multi}
Rozemberczki, B.; Allen, C.; and Sarkar, R. 2021.
\newblock Multi-scale attributed node embedding.
\newblock \emph{Journal of Complex Networks}, 9(2): cnab014.

\bibitem[{Rozemberczki and Sarkar(2020)}]{rozemberczki2020characteristic}
Rozemberczki, B.; and Sarkar, R. 2020.
\newblock Characteristic functions on graphs: Birds of a feather, from statistical descriptors to parametric models.
\newblock In \emph{Proceedings of the 29th ACM International Conference on Information and Knowledge Management}, 1325--1334.

\bibitem[{Scarselli et~al.(2008)Scarselli, Gori, Tsoi, Hagenbuchner, and Monfardini}]{scarselli2008graph}
Scarselli, F.; Gori, M.; Tsoi, A.~C.; Hagenbuchner, M.; and Monfardini, G. 2008.
\newblock The graph neural network model.
\newblock \emph{IEEE Transactions on Neural Networks}, 20(1): 61--80.

\bibitem[{Shah et~al.(2020)Shah, Dehmamy, Perra, Chinazzi, Barab{\'a}si, Vespignani, and Yu}]{shah2020finding}
Shah, C.; Dehmamy, N.; Perra, N.; Chinazzi, M.; Barab{\'a}si, A.-L.; Vespignani, A.; and Yu, R. 2020.
\newblock Finding patient zero: Learning contagion source with graph neural networks.
\newblock \emph{arXiv preprint arXiv:2006.11913}.

\bibitem[{Shah and Zaman(2011)}]{shah2011rumors}
Shah, D.; and Zaman, T. 2011.
\newblock Rumors in a network: Who's the culprit?
\newblock \emph{IEEE Transactions on Information Theory}, 57(8): 5163--5181.

\bibitem[{Srinivasan and Ribeiro(2019)}]{srinivasan2019equivalence}
Srinivasan, B.; and Ribeiro, B. 2019.
\newblock On the equivalence between positional node embeddings and structural graph representations.
\newblock In \emph{International Conference on Learning Representations}.

\bibitem[{Veli{\v{c}}kovi{\'c} et~al.(2017)Veli{\v{c}}kovi{\'c}, Cucurull, Casanova, Romero, Lio, and Bengio}]{velivckovic2017graph}
Veli{\v{c}}kovi{\'c}, P.; Cucurull, G.; Casanova, A.; Romero, A.; Lio, P.; and Bengio, Y. 2017.
\newblock Graph attention networks.
\newblock In \emph{International Conference on Learning Representations}.

\bibitem[{Wang, Jiang, and Zhao(2022)}]{wang2022invertible}
Wang, J.; Jiang, J.; and Zhao, L. 2022.
\newblock An Invertible Graph Diffusion Neural Network for Source Localization.
\newblock In \emph{Proceedings of the ACM Web Conference 2022}, 1058--1069.

\bibitem[{Wang et~al.(2017)Wang, Wang, Pei, and Ye}]{wang2017multiple}
Wang, Z.; Wang, C.; Pei, J.; and Ye, X. 2017.
\newblock Multiple source detection without knowing the underlying propagation model.
\newblock In \emph{Proceedings of the AAAI Conference on Artificial Intelligence}, volume~31.

\bibitem[{Wen et~al.(2017)Wen, Kveton, Valko, and Vaswani}]{wen2017online}
Wen, Z.; Kveton, B.; Valko, M.; and Vaswani, S. 2017.
\newblock Online influence maximization under independent cascade model with semi-bandit feedback.
\newblock \emph{Advances in Neural Information Processing Systems}, 30.

\bibitem[{Yang and Leskovec(2012)}]{yang2012defining}
Yang, J.; and Leskovec, J. 2012.
\newblock Defining and evaluating network communities based on ground-truth.
\newblock In \emph{Proceedings of the ACM SIGKDD Workshop on Mining Data Semantics}, 1--8.

\bibitem[{You, Ying, and Leskovec(2019)}]{you2019position}
You, J.; Ying, R.; and Leskovec, J. 2019.
\newblock Position-aware graph neural networks.
\newblock In \emph{International Conference on Machine Learning}, 7134--7143. PMLR.

\bibitem[{Zhou, Jagmohan, and Varshney(2019)}]{zhou2019generalized}
Zhou, H.; Jagmohan, A.; and Varshney, L.~R. 2019.
\newblock Generalized Jordan center: A source localization heuristic for noisy and incomplete observations.
\newblock In \emph{2019 IEEE Data Science Workshop (DSW)}, 243--247. IEEE.

\bibitem[{Zhu, Chen, and Ying(2017)}]{zhu2017catch}
Zhu, K.; Chen, Z.; and Ying, L. 2017.
\newblock Catch’em all: Locating multiple diffusion sources in networks with partial observations.
\newblock In \emph{Proceedings of the AAAI Conference on Artificial Intelligence}, volume~31.

\bibitem[{Zhu et~al.(2022)Zhu, Cheng, Gao, Wang, and Li}]{zhu2022locating}
Zhu, P.; Cheng, L.; Gao, C.; Wang, Z.; and Li, X. 2022.
\newblock Locating multi-sources in social networks with a low infection rate.
\newblock \emph{IEEE Transactions on Network Science and Engineering}, 9(3): 1853--1865.

\end{thebibliography}

\end{document}